\documentstyle[aps,prb,epsfig,ifthen,multicol]{revtex}

\newcommand{\wide}[2]{                                                                                             %
\end{multicols}                                                                                                                 %
\widetext                                                                                                                            %
\noindent                                                                                                                           %
\ifthenelse{\equal{#1}{t}}                                                                                              %
{}                                                                                                                                           %
{                                                                                                                                            %
\raisebox{0.1in}[0in][0.02in]{$\rule{3.575in}{0.002in}                                            %
\rule{0.002in}{0.08in}$}                                                                                                  %
}                                                                                                                                            %
#2                                                                                                                                         %
\ifthenelse{\equal{#1}{b}}                                                                                             %
{}                                                                                                                                           %
{                                                                                                                                            %
{\raisebox{-0.1in}[0in][0.02in]                                                                                       %
{\hspace{3.575in}$\rule{0.002in}{0.08in}                                                                   %
\rule[0.08in]{3.575in}{0.002in}$}                                                                                   %
}                                                                                                                                             %
}                                                                                                                                             %
\begin{multicols}{2}                                                                                                         %
\noindent                                                                                                                            %
}                                                                                                                                             %

\def  \bsig    {\mbox{\boldmath$\sigma$}}

\begin{document}
\draft

\title{A new mechanism of exchange interaction in ferromagnetic semiconductors}
\author{V.~K.~Dugaev$^{1,2,}$\cite{email}, V.~I.~Litvinov$^{3}$, J.~Barna\'s$^{4}$,
and M.~Vieira$^{1}$}
\address{$^1$Department of Electronics and Communications, Instituto
Superior de Engenharia de Lisboa,\\
Rua Conselheiro Emidio Navarro, 1949-014 Lisbon, Portugal\\
$^{2}$Institute for Problems of Materials Science, Vilde~5, 58001~Chernovtsy, Ukraine\\
$^{3}$WaveBand Corporation, 375 Van Ness Avenue, Suite 1105, Torrance,
California 90501\\
$^4$Department of Physics, Adam Mickiewicz University, ul. Umultowska 85,
61-614 Pozna\'n, Poland, and\\
Institute of Molecular Physics, Polish Academy of Sciences,
ul.~M.~Smoluchowskiego~17, 60-179~Pozna\'n, Poland}

\date{\today }
\maketitle

\begin{abstract}

We propose a new mechanism of indirect exchange interaction,
which can be responsible for the ferromagnetic
ordering in Mn-doped semiconductors (like GaMnAs)
at low carrier concentration. The mechanism is based on the interplay of the
hybridization of band states (conduction or valence) with localized
impurity (donor or acceptor) states and
the direct exchange interaction between localized spins and
the band states. The indirect exchange coupling between two impurities
occurs when the wavefunctions of the corresponding localized donor (acceptor)
states overlap. This coupling is independent of the free carrier concentration
and therefore may be responsible for ferromagnetic transition at low
or vanishing carrier concentration.
\end{abstract}
\pacs{75.50.Pp, 72.80.Ey, 75.50.Dd}

\begin{multicols}{2}

Recent progress in controlling spin-polarized electron transport
in metallic  magnetic structures, and successfull applications of such
systems in magnetoelectronics/sprintronics devices\cite{prinz}
renewed worldwide interest in ferromagnetic semiconductors
(FMS's). Experimental efforts are mainly focused on searching
for FMS's with the ferromagnetic transition occuring
above the room temperature. Current achievements, however, are still far from
this objective.
The most promising now seem to be III-V semiconductors doped with Mn atoms
(diluted magnetic semiconductors).
Recently, ferromagnetic transition at about 110~K was observed
in GaMnAs,\cite{ohno,beschoten} and this observation stimulated
further theoretical and experimental works aimed at understanding the origin and nature of
the transition.\cite{dietl1,koenig1}

It is commonly believed\cite{dietl,koenig} that the leading mechanism of
magnetic coupling between Mn$^{2+}$ ions, which leads to ferromagnetic phase
transition,
is the Rudermann-Kittel-Kasuya-Yosida (RKKY) indirect
exchange interaction {\it via} the
band carriers\cite{rkky} (holes in the case of GaMnAs).
However, this mechanism does not explain the occurence of ferromagnetism at
low or vanishing density of free carriers in GaMnAs.\cite{ohno}
Therefore, other mechanism(s) of magnetic coupling between Mn ions
is (are) expected to be responsible for ferromagnetism
at low carrier concentrations.
Inoue {\it et al}~\cite{inoue} proposed a mechanism based on the coupling
{\it via} resonant states due to Mn impurities, which can form for one
spin orientation (more specifically for minority spins)
close to the top of the valence band. This mechanism may lead to
ferromagnetism at low carrier concentration (also in the insulating phase).
The resonant states are formed from the $d$-states owing to the $p$-$d$
hybridization.
However, there is no clear evidence of such resonant states in GaMnAs.

Recently, another mechanism of
exchange interaction between Mn impurities was proposed, which
also may lead to ferromagnetism
at low carrier concentration.\cite{litvinov} The mechanism was based on
a modified version of the Bloembergen-Rowland type of exchange coupling,
in which virtual transitions between the valence band and the impurity acceptor band
were involved. This model explicitly takes into account
the presence of an impurity acceptor band, which in the case
of GaMnAs is formed near the top of the valence band.
In Ref.~[\onlinecite{litvinov}] the impurity band was approximated by
a very narrow band corresponding to large effective electron mass
and located above the valence band. However, position of the Mn acceptor
states in GaMnAs is not well established. It is generally believed
that the Mn impurities in GaMnAs act as acceptors ($d^5+$ hole).
At low impurity concentrations the holes are weakly bound to the
Mn impurities, leading effectively to an impurity acceptor state above the
valence band. When the Mn concentration increases, the impurity acceptor band
merges with the valence band.\cite{dietl1}
In that case the holes may propagate as quasi-free quasi-particles.
Owing to their exchange coupling to the localized moments of the Mn impurities,
they give rise to RKKY indirect exchange interaction.

In this paper we propose a new mechanism of indirect exchange coupling,
in which the localized impurity levels (donor or acceptor states) are
involved. The main point of the model is the interplay of
the following two interactions: (i)
direct exchange coupling of the band (conduction or valence)
states to localized spins, and (ii) hybridization of the localized donor
(acceptor) states
with the conduction (valence) band states.
The indirect exchange coupling between two localized spins is then
mediated {\it via} the localized impurity states.
This takes place when the wavefunctions of the
of the impurity states localized at different sites overlap.
Thus, the mechanism
may be relevant when the characteristic
size of the wavefunctions corresponding to the localized donor or acceptor states
is much larger than the characteristic
size of the wavefunctions corresponding to the localized spins ($d$-states).
If this is the case,
the direct coupling between two spins is negligible. When additionally
the RKKY mechanism is not applicable (e.g., the Fermi level is in the gap),
the indirect coupling {\it via} donor or acceptor states may be dominant.

Let us analyse now the exchange coupling between
magnetic impurities more quantitatively.
In the following we will consider the situation,
when a localized magnetic moment associated with each impurity
is due to the unfilled $d$-shell.
Apart from this, we assume that each impurity
gives rise to a localized impurity (donor or acceptor)
state that is spin degenerate. The localized spin is
coupled {\it via} direct exchange interaction
to the band states. On the other hand, the band states
are additionally hybridized with the localized donor or acceptor states.
Owing to these two interactions,
the information on the state of a localized spin
is transmitted to the localized donor (acceptor) state
(which is localized at the same lattice point as the corresponding spin).
To take this into account, we extract from the total Hamiltonian the
most important terms and write the relevant
effective model Hamiltonian in the form:
\begin{eqnarray}
\label{1}
H_{eff}=\sum _{\bf k} \psi ^\dag _{\bf k}\, \varepsilon _{\bf k}\, \psi _{\bf k}
+\sum _i \psi ^\dag _i\, \varepsilon _i\, \psi _i
\nonumber \\
+\frac1{\sqrt{N}}\sum _{{\bf k},i}g_{\bf k}\, \left(
\psi ^\dag _{{\bf k}\mu }\, \bsig _{\mu\nu}\cdot {\bf S}_i\, \psi _{i\nu}
+ H.c.\right),
\end{eqnarray}
where ${\bf S}_i$ is the spin localized at the point ${\bf R}_i$,
$N$ is the number of host atoms in the lattice, $\bsig $ are the Pauli matrices,
$\psi _{\bf k}$ and $\psi _i$ are the spinor operators corresponding
to the band electrons with the
energy spectrum $\varepsilon _{\bf k}\, $ and to
localized impurity states of energy  $\varepsilon _i$,
respectively. The first term describes the system of noninteracting
band electrons, while the second one corresponds to localized
impurity states (acceptor or donor states).
The third term describes that contribution to the effective
hybridization of the localized and band states,
which depends on the localized spin  ${\bf S}_i$. The other terms, which
do not contribute to the indirect coupling between two localized spins,
have been omitted in Eq.~(1).
In the third term $g_{\bf k}$ is an effective parameter that
describes ${\bf S}_i$-dependent contribution to the effective
hybridization of the band and donor (acceptor) states.
This constant includes the effects due to direct exchange
coupling between the localized moments ${\bf S}_i$ and the band states, and
can be derived {\it eg} by the perturbational methods. In the simplest
second order approximation one finds $g_{\bf k}\approx Jg_0/\varepsilon_d$,
where $J$ is the parameter of direct exchange
interaction between localized spins and
band states, $g_0$ is the bare hybridization parameter of the
band and localized states, and  $\varepsilon _d$ is the activation energy
of the donor (acceptor) states.

The total spin of the localized magnetic moment (${\bf S}_i$) and
of the delocalized (band states) and localized (donor or acceptor states)
electrons is conserved. This condition can be used to
eliminate some of the matrix elements of the effective Hamiltonian (1),
which do not conserve the total spin. In Eq.~(1) we assumed that all
the non-vanishing matrix elements are equal and denoted them by $g_{\bf k}$.

Hamiltonian (1) describes individual magnetic ions (impurities)
with the unfilled $d$-shell which is responsible for
the localized spin ${\bf S}_i$. These ions also give rise to
localized $s$ or $p$ states in the vicinity
of the valence or conduction bands. Such a situation takes place for
instance in GaMnAs,\cite{ohno} where the acceptor level associated
with a Mn impurity is formed near the top of the valence band.
To have exchange interaction between two
localized spins, the spin polarization of a localized state,
induced by one magnetic impurity, should be transmitted to
the place where the second magnetic impurity is located.
When these impurities, say indexed by $i$ and $j$, are not too far from each other,
the wavefunctions of the corresponding localized donor (acceptor) states,
$\psi _i$ and $\psi _j$, overlap. In that case, the direct hopping
between these localized states becomes important. In fact, it
is crucial for the indirect exchange interaction proposed in this paper.
Therefore, we must take it into account in the following considerations.
If we assume for simplicity, that the impurity states are of $s$
type, the corresponding term in the Hamiltonian may be written as
\begin{equation}
\label{2}
H_{hop}=\sum_{ij} \left( t_{ij} \psi ^\dag _i\, \psi _j+H.c.\right),
\end{equation}
where $t_{ij}$ denotes the hopping integral betwen states
$\psi _i$ and $\psi _j$.
When the distance between impurities, $R=\left| {\bf R}_i-{\bf R}_j\right| $,
is larger than the characteristic size
$r_0$ of the localized wavefunction,
one can assume $t_{ij}$ in the exponential form, $t_{ij}=A\exp (-R/r_0)$.
For $s$-type donor (acceptor) states with the activation energy
$\varepsilon _d$, one can estimate the parameters $A$ and $r_0$ as
$A\simeq \varepsilon _d$ and $r_0\simeq \hbar /(2m\varepsilon _d)^{1/2}$.

The total Hamiltonian of the system may be then written as
$H=H_{eff}+H_{hop}$. This is the basic Hamiltonian which includes all the
necessary interactions to describe the spin-spin coupling.
The exchange interaction between two spins, ${\bf S}_i$ and
${\bf S}_j$, can be calculated by the perturbation
method, using Eq.~(2) and the third term in Eq.~(1) as small perturbations.
The corresponding diagram for the coupling energy
is presented in Fig.~1, where the solid line represents the Green function
of band electrons, whereas the dashed lines correspond to the Green functions
of the $s$-type states localized at the points $i$ and $j$.
There is also another contribution which corresponds to the diagram similar to
that presented in Fig.~1, but with reversed orientation of the arrows.
The corresponding analytical expression for the interaction energy (including
contributions from both diagrams) has the form
\begin{equation}
\label{3}
E_{int}({\bf R}_i,{\bf R}_j)=w_{\alpha\beta}({\bf R}_i,{\bf R}_j)\,
S_{i\alpha }\, S_{j\beta} ,
\end{equation}

\begin{figure}
\psfig{file=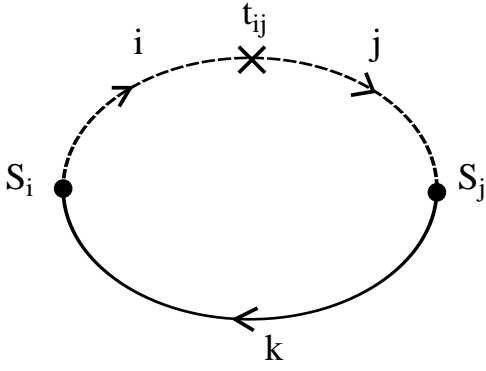,height=6cm}
\caption{Diagram contributing to  the indirect exchange interaction.}
\end{figure}

where
\begin{eqnarray}
\label{4}
w_{\alpha\beta}({\bf R}_i,{\bf R}_j)
=-\frac{ig^2\, t_{ij}}{n}\hskip3cm
\nonumber \\
\times\, {\rm Tr}\int \frac{d\varepsilon }{2\pi }\;
\left[ \sigma _\alpha \; G_i(\varepsilon )\; G_j(\varepsilon )\;
\sigma _\beta \; G(\varepsilon ;\, {\bf R}_j,{\bf R}_i)
\right. \nonumber \\
\left.
+\, \sigma _\alpha \; G(\varepsilon ;\, {\bf R}_i,{\bf R}_j)\;
\sigma _\beta \; G_j(\varepsilon )\; G_i(\varepsilon )\right] .
\end{eqnarray}
Here, $n=N/\Omega $ is the density of host atoms (with $\Omega $
denoting the crystal volume), whereas
$G(\varepsilon ;\, {\bf R}_i,{\bf R}_j)$ and $G_i(\varepsilon )$ are the
Green functions of the band and localized electrons, respectively.
For simplicity, the effective hybridization parameters
$g_{\bf k}$ are taken constant,
$g_{\bf k}=g$. In the case of a simple parabolic band,
$\varepsilon _{\bf k}=\hbar ^2k^2/{2m}$,
the corresponding Green functions take the form
\begin{eqnarray}
\label{5}
G(\varepsilon ;\, {\bf R}_i,{\bf R}_j)
=\int \frac{d^3{\bf k}}{(2\pi )^3}\;
\frac{\exp \left[ i{\bf k}\cdot ({\bf R}_i-{\bf R}_j)\right] }
{\varepsilon -\varepsilon _{\bf k}+i\delta \, {\rm sgn}\, (\varepsilon -\mu )}
\nonumber \\
=-\frac{m\, \exp \, (i\left|{\bf R}_i-{\bf R}_j\right|\, \kappa )}
{2\pi \hbar ^2\left|{\bf R}_i-{\bf R}_j\right|\, \kappa } .
\end{eqnarray}
On the other hand, for the localized states
$\varepsilon _i=\varepsilon _j=\varepsilon _0$
one finds
\begin{equation}
\label{6}
G_i({\varepsilon })=\left[ \varepsilon -\varepsilon _0
+i\delta \, {\rm sgn}\, (\varepsilon -\mu)\right] ^{-1}.
\end{equation}
In the above equations $\mu $ stands for the chemical potential and
$\kappa$ is defined as $\kappa =(2m\varepsilon )^{1/2}/\hbar$. The integration
over $\varepsilon $ in Eq.~(4) is to be carried out along the real axis with
an infinitezimaly small shift ($-i\delta $)
into the negative imaginary part of the complex plane
for $\varepsilon <\mu $, and with a small shift ($i\delta $)
into the positive imaginary part
of the complex plane for $\varepsilon >\mu $.

Let us consider now in more details the case of $\varepsilon _0<0$ and
$\varepsilon _0<\mu <0$. This situation corresponds to the localized
donor level ($\varepsilon _d=-\varepsilon _0$) filled with electron, and an
empty conduction band at $T=0$.
In the limit $T\rightarrow 0$, the chemical potential $\mu $  can be taken near
the bottom of the conduction band, $\mu =0^-$.
Using Eqs.~(4) to (6), we can calculate the integral over $\varepsilon $ by
closing the integration path in the upper half-plane of the
complex variable
$\varepsilon $ like presented in Fig.~2, where the thick line
shows the cut in the complex plane related to $\sqrt{\varepsilon }$.
Only the second-order pole at
$\varepsilon =\varepsilon _0$ contributes to the integral.
After calculating Eq.~(4), we arrive at the following expression for the
coupling energy:
\begin{equation}
\label{7}
w_{\alpha\beta}(R)=-\delta _{\alpha\beta}\;
\frac{2g^2\, t_{ij}\, m^2}{\pi \hbar ^4 \kappa _0\, n}\;
{\rm e}^{-R\, \kappa _0}
,\end{equation}
where $R=\left| {\bf R}_i-{\bf R}_j\right| $ and
$\kappa _0=(2m\left| \varepsilon _0\right| )^{1/2}/\hbar $.

The indirect exchange coupling described by Eq.~(7)
originates from the magnetic polarization of the lattice via the
impurity states and is ferromagnetic in sign.
Contrary to the RKKY coupling, the interaction
described by  Eqs.~(3) and (7) does not depend on the free carrier
concentration, and therefore may exist even in non-degenerate
semiconductors at $T=0$.

Taking the approximate formula for $t_{ij}$, as described after Eq.~(2) and
valid in the case
when the distance between impurities
is larger than the characteristic size of the localized wavefunctions,
we obtain the interaction range function in the form
\begin{equation}
\label{8}
w_{\alpha\beta}(R)\simeq -\delta _{\alpha\beta}\;
\frac{2g^2\, \left| \varepsilon _0\right| m^2}{\pi \hbar ^4 \kappa _0\, n}\;
{\rm e}^{-2R\, \kappa _0}.
\end{equation}
Let us estimate now magnitude of the interaction.
For $R\kappa _0\simeq 1$ we obtain
$\left| E_{int}\right| \sim (g/\left| \varepsilon _0\right| )^2
\left| \varepsilon _0\right| $. Thus, one can expect maximum of the
interaction at $\left| g/\varepsilon _0\right| \sim 1$, when
$\left| E_{int}\right| _{max}\sim \left| \varepsilon _0\right| $.
For $\left| g/\varepsilon _0\right| \simeq 0.3$ and
for $\left| \varepsilon _0\right| \simeq 100$~meV, we get
$\left| E_{int}\right| \simeq 100$~K.

The above calculation of the exchange interaction (Eqs.~7 and 8)
applies to the situation, where the donor
level is below the conduction band. The calculation is also valid when
the acceptor level is located above the valence band, simply
by replacing electrons with holes.

\begin{figure}
\psfig{file=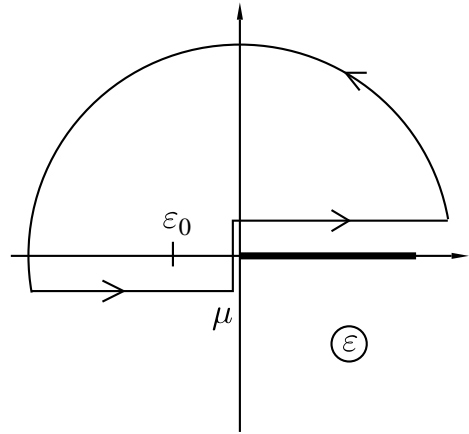,height=8 cm}
\caption{Path of the integration in Eq.~(4).}
\end{figure}

However, the structure of the valence bands in zinc-blende
semiconductors is more complex and cannot be described by
a simple parabolic model. The proposed mechanism, however,
is still valid in this case, but the
interaction should have a form which is more complex than that
describeed by Eq.~(7).

If an acceptor level forms a resonance in the valence band,
the mechanism presented in this letter is working as well,
provided  the Fermi level lies
between the level $\varepsilon _0$ and the top of valence band.
In that case the acceptor level behaves like a pseudo-donor, and the
virtual transitions of electrons between the level and the valence band are
responsible for the spin-spin coupling. However,
the usual RKKY interaction {\it via} free holes
can dominate the exchange spin-spin interaction in this case.

It should be mentioned at this point,
that the ferromagnetism induced by resonance
impurity-band mixing in degenerate semiconductors has been
considered long time ago by Abrikosov.\cite{abrikosov} Magnetic
moments and ferromagnetism appear then,
when the impurity level is situated close to the Fermi energy.
Contrary to this model, in our case
the magnetic moments (${\bf S}_i$) are formed by deep
$d$-states of the impurities, and their mixing to the band states
{\it via} direct exchange coupling gives rise to indirect
exchange interaction between localized spins, and consequently
to ferromagnetism.

In conclusion, we proposed in this Letter
a new mechanism of exchange interaction of
magnetic impurities in semiconductors.
The mechanism takes into account the fact that
the $d$-shell spin ${\bf S}_i$ is located at the same
point as the shallow localized impurity state $\psi _i$ (donor or acceptor).
The main point of the mechanism is the interplay of
the direct exchange interaction of the band states and localized spins,
and the hybridization of the band states with the localized donor
or acceptor states. The information on the spin state of the
impurity is then transmitted from one impurity to the other
owing to the overlap of the corresponding wavefunctions
of the localized states.

The mechanism favors ferromagnetic ordering of the impurity spins
and the coupling energy is independent of the free carrier density.
Therefore, this mechanism may play an important (or even crucial) role
at low carrier concentrations,
and may be responsible for
the ferromagnetism observed in such systems like for instance
GaMnAs alloy semiconductors.

V.D. thanks V.~R.~Vieira and P.~D.~Sacramento for comments.
This work is partially supported by the Polish State Committee for Scientific
Research through the Project No.~5~03B~091~20, NATO Linkage
Grant No.~977615, and NATO Science fellowship CP(UN)06/B/2001/PO.

\end{multicols}

\end{document}